\let\latexarabic\arabic
\let\latexdocument\document
\let\latexenddocument\enddocument

\RequirePackage[thmmarks]{ntheorem}
\makeatletter
\renewtheoremstyle{plain} 
  {\item[\hskip\labelsep \theorem@headerfont ##1\ \textup{##2}\theorem@separator]} 
  {\item[\hskip\labelsep \theorem@headerfont ##1\ \textup{##2}\ (##3)\theorem@separator]}
\makeatother

\documentclass[article]{biomka}

\let\document\latexdocument
\let\enddocument\latexenddocument
\AtEndDocument{\printhistory}
\let\arabic\latexarabic
\def\rm{}

\usepackage{amsmath,amssymb}
\usepackage{booktabs}
\usepackage[T1]{fontenc}

\usepackage{times}
\usepackage{bm}
\usepackage{natbib}

\usepackage[plain,noend]{algorithm2e}

\usepackage{color}
\usepackage{multirow}

\makeatletter
\renewcommand{\algocf@captiontext}[2]{\quad #1\algocf@typo. \AlCapFnt{}#2} 
\def\@algocf@capt@plain{top}
\renewcommand{\algocf@makecaption}[2]{%
  \addtolength{\hsize}{\algomargin}%
  \sbox\@tempboxa{\algocf@captiontext{#1}{#2}}%
  \ifdim\wd\@tempboxa >\hsize
    \hskip .5\algomargin%
    \parbox[t]{\hsize}{\algocf@captiontext{#1}{#2}}
  \else%
    \global\@minipagefalse%
    \hbox to\hsize{\box\@tempboxa}
  \fi%
  \addtolength{\hsize}{-\algomargin}%
}
\makeatother

\makeatletter
\ifx\selectfont\undefined
   \def\Operator#1{\mathop{\rm#1}\nolimits}
   \def\Roman#1{{\rm#1}}
\else
   \def\Operator#1{\mathop{\operator@font#1}\nolimits}
   \let\Roman=\mathrm
\fi
\makeatother

\newcommand{\graf}{\mathcal{G}}

\newcommand{\bound}{\Operator{bd}}
\newcommand{\clos}{\Operator{cl}}

\newcommand{\transp}{^{\top}}

\newcommand{\trace}{\Operator{tr}}

\newcommand{\cd}{\,|\,}

\newcommand{\rank}{\Operator{rank}}

\newcommand{\seteq}{\leftarrow}

\def\inv{^{-1}}
\def\tildeK{(S_{cc})\inv}
\def\tildeS{S_{cc}}

\newcommand{\norm}[1]{\ensuremath{\lVert#1\rVert}}

\newcommand{\mnorm}[1]{\ensuremath{\lVert|#1|\rVert}}

\newcommand{\paragraph}{\emph}
\def\pkg#1{\textsc{#1}}

\newcommand{\col}{\text{col}}

\begin{document}




\markboth{S. H{\o}jsgaard and S. Lauritzen}{Algorithms for estimation in Gaussian graphical models}

\title{On some algorithms for estimation \\in Gaussian graphical models}

\author{S. H\o jsgaard}
\affil{Department of Mathematical Sciences, Aalborg University\\ Skjernvej 4A, DK-9220 Aalborg, Denmark \email{sorenh@math.aau.dk}}

\author{S. Lauritzen}
\affil{Department of Mathematical Sciences, University of Copenhagen\\ Universitetsparken 5, DK-2100 Copemhagen, Denmark
\email{lauritzen@math.ku.dk}}

\maketitle

\begin{abstract}
 In Gaussian graphical models, the likelihood equations must
  typically be solved iteratively. We investigate two algorithms: A version of iterative proportional scaling 
  which avoids  inversion of large matrices, and an 
  algorithm based on convex duality and operating on the covariance matrix by neighbourhood coordinate descent,  corresponding to the graphical lasso with zero penalty. For large, sparse graphs, the iterative proportional scaling algorithm appears feasible and has simple convergence properties. The algorithm based on neighbourhood coordinate descent is extremely fast and less dependent on sparsity, but needs a positive definite starting value to converge. We give an algorithm for finding such a starting value for graphs with low colouring number. As a consequence, we also  obtain a simplified proof for existence of the maximum likelihood estimator in such cases. 
\end{abstract}

\section{Introduction}
\label{sec:introduction}
Maximum likelihood estimation in Gaussian graphical models can be carried out via generic optimization
algorithms, Newton--Raphson iteration, iterative proportional scaling, other alternating algorithms \citep{speed:kiiveri:86}, and algorithms exploiting duality and operating on the covariance matrix, such as the algorithm of \cite{wermuth:scheidt:77} and neighbourhood coordinate descent  \citep[p.\ 631 ff.]{hastie:etal:16}. Neighbourhood coordinate descent may be seen as an instance of the
graphical lasso with zero penalty \citep{yuan:lin:07,banerjee:etal:08,friedman:etal:08}, or as a special instance of the GOLAZO algorithm \citep{lauritzen:zwiernik:22}. 

Algorithms based on duality need a positive definite starting value to guarantee convergence and this may be challenging when the number of variables exceeds the number of observations. In addition, the concentration matrix returned by these algorithms after finitely many steps may not have exact zero values for entries corresponding to non-edges, and additional procedures are neccessary to ensure this.

In contrast,  iterative proportional scaling is provably convergent  from the identity matrix as a starting value when the maximum likelihood estimator exists and it satisfies the model restrictions at all times; but it may be slow in high dimensions. 

The main contributions of this article are the following: We present a version of the iterative proportional scaling algorithm that updates the covariance and concentration matrices simultaneously and works on edges rather than cliques, so identification of cliques becomes unnecessary and inversion of large matrices is avoided. Further, we present a version of neighbourhood coordinate descent which generates a positive definite starting value if the graph is sufficiently sparse and is guaranteed to output a positive definite concentration matrix.

Alternative ways of speeding up the iterative proportional scaling algorithm 
typically involves special methods for calculating appropriate marginals, for example using probability propagation as described for the analogous discrete case in \cite{jirousek:preucil:95}.  Approaches along these lines have been used by 
\cite{hara:takemura:10} and \cite{xu:etal:11,xu:etal:12}.  \cite{xu:etal:15} give a thorough survey and comparison of the methods and also show how to speed up the scaling algorithms by partitioning of the cliques and localized updating. 
The methods investigated in this article are based on simple and general matrix manipulations and avoid setting up   more involved computational structures. 




\section{Likelihood equations for Gaussian graphical models}
\label{sec:likel-equat-ggms}
 Let
$X=(X_v, v\in V)$ be a $d$ dimensional random vector, i.e.\
$|V|=d$, normally distributed with mean zero and covariance matrix $\Sigma$. The focus
is on the pattern of zeros in the inverse covariance matrix, i.e. in
the concentration matrix $K=\Sigma\inv$. 
If $K_{uv}=0$ then $X_u$ and $X_v$ are conditionally independent given
$X_{V\setminus \{u,v \}}$. The pattern of zeros in $K$ may be
represented by an undirected graph $\graf=(V,E)$ with vertices $V$ and
edges $E$. A Gaussian graphical model is then defined by demanding $K_{uv}=0$ unless there is an edge $uv\in E$.
For further details, we refer to \citet[Ch.\ 4]{lauritzen:96}.

Let $\graf=(V,E)$ be a simple, undirected graph and let $S$ denote the empirical covariance
matrix obtained from a sample $X^1=x^1,\ldots, X^n=x^n$.  The maximum likelihood estimate $\hat\Sigma$ of  the covariance matrix $\Sigma$ in an undirected Gaussian
graphical model is the unique  solution to the system of equations \citep[p.\ 133]{lauritzen:96}:
\begin{eqnarray}  
  \hat\Sigma_{vv}-S_{vv}&=& 0\;\mbox{ for all $v\in V$,}\label{eq:diag}\\
  \hat\Sigma_{uv}-S_{uv}&=& 0\;\mbox{ for all $uv\in E$,}\label{eq:edge}\\
\hat K_{uv}=(\hat \Sigma^{-1})_{uv}&=& 0\;\mbox{ for all $uv \not \in  E$,}\label{eq:slackness}
\end{eqnarray}
provided such a solution exists.  If we for a matrix $A$  let $A(\graf)$ be the  matrix obtained  by replacing entries $A_{uv}$ with zero  for $uv\notin E$, we may reexpress (\ref{eq:diag}) and (\ref{eq:edge}) as 
\begin{equation}\label{eq:alt_edge} \hat\Sigma(\graf)=S(\graf).\end{equation} 

We consider two types of algorithm for solving these equations. One, iterative proportional scaling, always   obeys  the constraints in (\ref{eq:slackness}) and attempts to make (\ref{eq:alt_edge}) satisfied by successive iterations.
The other type of algorithm, neighbourhood coordinate descent, always obeys  the constraints in (\ref{eq:alt_edge}) but attempts through successive iterations to satisfy (\ref{eq:slackness}). 

\section{Iterative proportional scaling}
\subsection{Computational issues of updating}


Iterative proportional scaling cycles through  subsets
$c\subseteq V$ of variables which are complete in $\graf$, i.e.\ all elements of $c$ are mutual neighbours in the graph. The current estimate of $\Sigma$ is  updated by keeping 
parameters of the conditional distribution $X_{V\setminus c}\cd X_c$
fixed, whereas  parameters of the marginal distribution of $X_c$
are updated to maximize the objective function under that
restriction. The updates have the form
\begin{equation}\label{eq:scaling}f(x; \Sigma)\; \leftarrow \;f(x) \frac{f(x_c; S_{cc})}{f(x_c; \Sigma_{cc})},\end{equation}
where $S_{cc}$ and $\Sigma_{cc}$  here and in the following indicate the corresponding marginals of the empirical covariance matrix and of the current value of $\Sigma$; hence 
the densities are scaled proportionally, whence the name of the algorithm.
If the maximum likelihood estimate exists, the algorithm is convergent when started in a point satisfying the model restrictions, for example when started at $K=I$    \citep[Thm.\ 5.4]{lauritzen:96}.

Let $c\subseteq V$ and $a=V\setminus c$, where $c$ is a complete subset
of $V$ in $\graf$.
The update for  $c$ of the concentration matrix $K$ takes the form \citep[p.\ 134]{lauritzen:96}
\begin{equation}\label{eq:kupdate1}
  K_{cc} \seteq \tildeK + L,
\end{equation}
whereas $K_{ac}, K_{aa}, K_{ca}$ are unchanged.
There are two alternatives for calculating $L$:
\begin{eqnarray}
  L &=&K_{ca}(K_{aa})\inv K_{ac}   \label{eq:Lupd1} \\
    &=& K_{cc}-(\Sigma_{cc})\inv.  \label{eq:Lupd2}
\end{eqnarray}
Calculating $L$ as in (\ref{eq:Lupd1}) gives what is referred to in this
paper as the \emph{concentration version} of the algorithm. Expression (\ref{eq:Lupd1})
has the advantage that
$\Sigma= K\inv$ is not needed, so inversion of $K$ is avoided and
$\Sigma$ need not be stored. This is efficient if $a$ is small and
$c$ is large.

Calculating $L$ as in (\ref{eq:Lupd2}) gives what is referred to in this
paper as the \emph{covariance version} of the algorithm.
Expression (\ref{eq:Lupd2}) has the advantage that computation of
$(K_{aa})\inv$ is not needed and this matrix inversion could be expensive if $a$ is
large. On the other hand, $\Sigma$ needs to be stored and calculated. 
Luckily it is possible to update 
$\Sigma$ along with  $K$, avoiding repeated and time consuming matrix inversions. 
This makes expression (\ref{eq:Lupd2})
feasible to use in practice and speeds up the computation considerably.


\subsection{Updating $\Sigma$}
\label{sec:updat-sigma-with}

%
The updated version $\tilde \Sigma$ of $\Sigma$ can be calculated as
\begin{equation}\label{eq:sigmaupdate}\tilde \Sigma = \begin{pmatrix}S_{cc} &S_{cc}(\Sigma_{cc})^{-1}\Sigma_{ca}\\ \Sigma_{ac}(\Sigma_{cc})^{-1}S_{cc}&
\;\Sigma_{aa}-\Sigma_{ac}H\Sigma_{ca}\;\end{pmatrix},\end{equation}
where 
$H$ is equal to
 \begin{equation}\label{eq:harville_red} 
 H=(\Sigma_{cc})\inv -(\Sigma_{cc})\inv\tildeS(\Sigma_{cc})\inv\end{equation}
and thus inversion of  
 $K$ is avoided. 
To see this is correct, we may establish that
\begin{equation}\label{eq:inv_relation}\begin{pmatrix}S_{cc} &S_{cc}(\Sigma_{cc})^{-1}\Sigma_{ca}\\ \Sigma_{ac}(\Sigma_{cc})^{-1}S_{cc}&
\;\Sigma_{aa}-\Sigma_{ac}H\Sigma_{ca}\;\end{pmatrix}=\begin{pmatrix}(S_{cc})\inv +K_{ca}(K_{aa})\inv K_{ac}  &\;K_{ca}\;\\K_{ac}&K_{aa}\end{pmatrix}^{-1},\end{equation}
which follows by direct matrix multiplication using the identities
$$(\Sigma_{cc})\inv\Sigma_{ca}= -K_{ca}(K_{aa})\inv ,\quad
(K_{aa})\inv = \Sigma_{aa}-\Sigma_{ac}(\Sigma_{cc})\inv \Sigma_{ca}.$$

The update of $\Sigma$ as in (\ref{eq:sigmaupdate}) is also given as formula (19) in \cite{speed:kiiveri:86}, on page 185 of \cite{whittaker:90}, and as formula (2) of \cite{xu:etal:15}, albeit in varying notations.



If the difference between $\Sigma_{cc}$ and $S_{cc}$ is small, the corresponding update may be ignored altogether, see further in Section~\ref{sec:precision-algorithms}.  

There are  two natural choices for the system of complete sets $c$: the set $\mathcal C$ of cliques of $\graf$; or the set of edges $E$. Both choices are compared in our experiments, but we focus on  the set of edges $E$, as this avoids the NP-complete task of determining the cliques of the graph.


\subsection{Updating the  likelihood function}
\label{sec:updat-value-likel}If the likelihood function should be monitored, this can also be updated in a simple fashion.
Consider again a sample  $X^1=x^1, \dots, X^n=x^n$ where
$X^\nu \sim N_d(0, \Sigma)$ and  
let $S$ denote the sample covariance matrix. The log-likelihood function (ignoring additive constants) is
\begin{equation}\label{eq:loglik}
  l (K) = \frac n 2 \log \det(K) - \frac n 2 \trace(KS).
\end{equation}
From (\ref{eq:inv_relation}) we get
$$ \det \tilde K =\det\{\tilde K_{cc}-K_{ca}(K_{aa})^{-1}K_{ac}\}\det K_{aa}= \det (S_{cc})^{-1}\det K_{aa}.$$
But 
$\det K = \det (\Sigma_{cc})^{-1} \det K_{aa}$. Hence, if we let $A= (\Sigma_{cc})^{-1}S_{cc}$ we have
$$\log \det \tilde K = \log\det K - \log  \det A.$$
If we let $\Delta_{cc}$ denote the difference between the updated $\tilde K_{cc}$
and old $K_{cc}$ i.e.\
\begin{equation}\label{eq:delta1}
  \Delta_{cc}=\{ \tildeK +L\}- K_{cc}=\tildeK-(\Sigma_{cc})\inv,
\end{equation}
we have
for the trace term from  (\ref{eq:delta1})
\begin{eqnarray*}\trace(\tilde KS)&=&\trace(KS)+\trace(\Delta_{cc} S_{cc})\\&=&\trace(KS)+|c|-\trace\{(\Sigma_{cc})^{-1} S_{cc}\}=\trace(KS)+|c|-\trace(A),\end{eqnarray*}
where $|c|=\trace\{(S_{cc})^{-1}S_{cc}\}$ is the size of  $c$.
We thus get the expression
$$
  \ell (\tilde K)=
  \ell(K) -\frac n 2|c|-\frac n 2\log \det A+\frac n 2\trace (A).
$$
Here the matrix $A$ has dimension $|c|\times|c|$ so the adjustment is easily calculated if $c$ is small.

\subsection{Convergence issues}
\label{sec:precision-algorithms}

Convergence of  the algorithms can be assessed by investigating whether the likelihood
equations (\ref{eq:alt_edge}) are satisfied within a
small numerical threshold since (\ref{eq:slackness}) remains exactly satisfied at all times because the starting value satisfies the model restrictions and only values along edges are changed.  This requires
that $\Sigma$ is available and we may then express the deviation as
$\norm{\Sigma(\graf)-S(\graf)}_\infty$, where 
$$\norm{A}_\infty=\max_{uv}|A_{uv}|$$ is the maximum absolute deviation norm. 
The gradient of the log-likelihood function (\ref{eq:loglik}) is equal to
$$\nabla_K\ell(K)=\frac{n}{2}\{\Sigma(\graf)-S(\graf)\}$$ so it seems appropriate to continue the iteration until the size of the gradient is small, hence until 
$$\norm{\Sigma(\graf)-S(\graf)}_\infty\leq 2\varepsilon/n$$
where $\varepsilon$ is a small threshold. 

Commonly used,  but less stringent, convergence criteria monitor whether changes in the log-likelihood or changes in
parameter values between succesive iterations are small.  However, one may
find that such changes only indicate that the algorithm slows down  even though the likelihood
equations are far from being satisfied and the value of the likelihood function may be far from its maximum. 
\subsection{Computational savings}
\label{sec:effort_ips}
To achieve a further speedup of the algorithm we may  for each  $c$ check whether $\norm{\Sigma_{cc}-S_{cc}}_\infty<2\varepsilon/n$ before the update is executed. If  this is the case we may ignore the update as this will be time consuming but ineffective; this allows the algorithm to move faster towards the limit when the algorithm is close but not close enough. If the algorithm is terminated when no local updates are needed,   the likelihood equations (\ref{eq:alt_edge}) are still satisfied within the same threshold, since
$$\norm{\Sigma(\graf)-S(\graf)}_\infty=\max_{c\in \mathcal A}\,\norm{\Sigma_{cc}-S_{cc}}_\infty,$$
where $\mathcal A$ is the chosen system of  subsets.
\section{Algorithms based on convex duality}
\subsection{The optimization problem and its dual}\label{sec:dual}
The problem of maximizing the log-likelihood function, to be referred to as the \emph{primal problem}, may be formulated as follows, where as before $S$ denotes the empirical covariance matrix and we have ignored the multiplicative constant $n/2$:
\begin{equation}
\label{eq:primal}
\begin{aligned}
& \underset{K}{\text{maximize}}
& &\ell(K)=\log\det(K) - \trace(KS)  \\
& \text{subject to}
&& K\in\mathbb{S}^{d\times d}_{\succ}(\graf), \\
\end{aligned}
\end{equation}
where $\mathbb{S}^{d\times d}_{\succ}(\graf)$ denotes the set of positive definite matrices $K$ with $K_{uv}=0$ for all $uv\notin E(\graf)$.
This is a convex optimization problem with a unique solution if and only if the maximum likelihood estimate exists. We shall for the moment assume that $S$ is positive definite,  but later identify necessary modifications  for a more general case.
To exploit convex duality \citep[Ch.\ 5]{boyd:vandenberghe:04}, we consider the Lagrangian
$$\mathcal{L}(\Lambda, K)=\log\det(K) -\trace(KS)-\trace(K\Lambda),$$
where $\Lambda$ is a symmetric matrix satisfying $\Lambda_{uv}=0$ for all $uv\in E$ and $\Lambda_{uu}=0$ for all $u\in V. $ We now get the dual  function
$$g(\Lambda)=\sup_{K}\mathcal{L}(\Lambda,K)=- \log\det(S+\Lambda)-d.$$
  Since  for $K\in \mathbb{S}^{d\times d}(\graf)$ we have $\mathcal{L}(\Lambda,K)=\ell(K)$, the dual function yields an upper bound on $\ell(K)$ which we now wish to minimize.
Letting $\Sigma=S+\Lambda$ yields the dual optimization problem as
\begin{equation*}
\begin{aligned}
& \underset{\Sigma}{\text{minimize}}
& &g(\Lambda)=-\log\det(\Sigma) - d  \\
& \text{subject to}
&& \Sigma \in\mathbb{S}_{\succ}^{d\times d}, \quad  \Sigma(\graf)=S(\graf)\\ 
\end{aligned}
\end{equation*}
or, equivalently, 
\begin{equation}
\label{eq:dual2}
\begin{aligned}
& \underset{\Sigma}{\text{maximize}}
& &\det(\Sigma) \\
& \text{subject to}
&& \Sigma \in\mathbb{S}_{\succ}^{d\times d}, \quad\Sigma(\graf)=S(\graf).
\end{aligned}
\end{equation}
A feasible point for (\ref{eq:dual2}) is often referred to as a \emph{positive definite completion} \citep{grone:etal:84} of the partial matrix $$\Sigma_\graf=\{\Sigma_{uu},u\in V; \Sigma_{uv}, uv\in E\}.$$ The maximum likelihood estimator of $\Sigma$ exists if and only if there is such a feasible point. It then holds that $\Sigma$ is the unique optimizer of (\ref{eq:dual2}) if and only if $K=\Sigma^{-1}$ is the unique optimizer of (\ref{eq:primal}).

\subsection{Solving the dual problem}
For a single vertex $u\in V$ we  let $c= V\setminus \{u\}$,  $b=\bound(u)$, and $r=\clos(u)^c$; so the variables $X$ may be  partitioned as 
$$X=(X_{\clos(u)^c},X_{\bound(u)}, X_u)\transp = (X_{r},X_{b}, X_{u})\transp=  (X_{c},X_{u})\transp.$$ Similarly, we write the covariance matrix $\Sigma$ in block form as
$$\Sigma= \begin{pmatrix}\Sigma_{rr}&\Sigma_{rb}&\Sigma_{ru}\\
\Sigma_{br}&\Sigma_{bb}&\Sigma_{bu}\\
\Sigma_{ur}&\Sigma_{ub}&\Sigma_{uu}
\end{pmatrix}= \begin{pmatrix}\Sigma_{cc}&\Sigma_{cu}\\
\Sigma_{uc}&\Sigma_{uu}
\end{pmatrix}.$$
Using the Schur complement $(\Sigma/\Sigma_{cc})$ we may express the determinant as
\begin{equation}\label{eq:factordet}\det \Sigma =\det \Sigma_{cc} \det(\Sigma/\Sigma_{cc})=\det \Sigma_{cc} \left\{\Sigma_{uu}-\Sigma_{uc}(\Sigma_{cc})^{-1}\Sigma_{cu}\right\}.\end{equation}
We should now keep  $\Sigma_{cc}$  fixed and maximize the  Schur complement over feasible values of $\Sigma_{uc}$. But the Schur complement is the residual variance after regressing $X_u$ on the remaining variables. Since $\Sigma_{bu}$ is fixed, the residual variance is maximized by ignoring the variables in $r$, which leads to changing the entries corresponding to non-neighbours $r$ of $u$ as  
\begin{equation}\tilde \Sigma_{ru}= \Sigma_{rb}(\Sigma_{bb})^{-1}S_{bu}=\Sigma_{rb}(\Sigma_{bb})^{-1}\Sigma_{bu}=\Sigma_{rb}\beta_{bu}, \label{eq:update}\end{equation}whereas all other entries of $\Sigma$ are unchanged. Here and in the following we have let $$\beta_{ub}=\beta_{bu}\transp=\Sigma_{ub}(\Sigma_{bb})^{-1}$$ denote the vector of regression coefficients for regressing $X_u$ on $X_b$. To get the first expression in (\ref{eq:update}) we have used that $\Sigma_{uv}=S_{uv}$ for all $uv\in E$. 
If $b=\emptyset$ we simply update by letting $\tilde\Sigma_{ru}=0$. 
We introduce the notation $D_u$ for this operation so that $\tilde\Sigma=D_u\Sigma$. 

Comparing to the update (\ref{eq:scaling}) for iterative proportional scaling, we here keep the marginal distribution of variables in $c$ fixed, whereas the conditional distribution of $X_u$ given the remaining variables is updated to reflect the conditional independence of $X_u$ and $X_r$, given the boundary variables $X_b$. 

 If the graph is sparse, the boundary $b$ is typically a small set and the expression (\ref{eq:update}) therefore avoids  inversion of large matrices.  Simple manipulations yield  the updated value of the Schur complement  
\begin{equation}\label{eq:schur}(\tilde\Sigma/\tilde\Sigma_{cc})=\tilde \Sigma_{uu}-\tilde\Sigma_{uc}(\tilde \Sigma_{cc})^{-1}\tilde\Sigma_{cu}=S_{uu}-S_{ub}(\Sigma_{bb})^{-1}S_{bu}=S_{uu}-S_{ub}\beta_{bu},\end{equation}
reflecting that after the update, 
the regression on the variables in $c$ is identical to the regression on the variables in $b$ only.

The optimization step following equation (\ref{eq:factordet}) is also the optimization step in the  GOLAZO algorithm \citep{lauritzen:zwiernik:22} when specialized to estimation in Gaussian graphical models. The GOLAZO algorithm solves this by quadratic programming, but in the  case considered, the optimization is simple and explicit, as given in (\ref{eq:update}).  Also, it is easy to verify that the update step  in (\ref{eq:update}) is identical to the update step used in \cite{hastie:etal:16}, Section 17.3.1; we refrain from giving the details as this is only a matter of comparing notations.  We shall in the following refer to the algorithm as \emph{neighbourhood coordinate descent}.

Another way to solve the dual problem visits non-edges $uv\notin E$ in turn and factor the determinant as 
$$\det(\Sigma)= \det\Sigma_{AA} \det \left\{\Sigma_{BB}-\Sigma_{BA}(\Sigma_{AA})^{-1}\Sigma_{AB}\right\},$$ where $A=V\setminus\{u,v\}$ and $B=\{u,v\}$.   One may then maximize the second factor in $\Sigma_{AB}$ keeping all other elements  of $\Sigma $ fixed. This can easily be done explicitly; see for example \cite{uhler:19}. This algorithm was implemented by \cite{wermuth:scheidt:77} but we shall not investigate it further, as it will be  slow when graphs are large and sparse, since then the number of non-edges  becomes huge. 
\subsection{Monitoring convergence}\label{sec:mon_conv}
Neighbourhood coordinate descent is implemented in the R package \pkg{ggm} \citep{marchetti:etal:20}. Since  neighbourhood coordinate descent is identical to the graphical lasso algorithm with zero penalty, see p.\ 637 in \cite{hastie:etal:16}, it is in effect also implemented in the  \pkg{glasso} package
\citep{glasso:19}. In both of these implementations, convergence is monitored by changes in $\Sigma$ after a full round of updating, so the algorithm is halted when consequtive values of $\Sigma$ are identical up to a given tolerance. As mentioned earlier, we find this to be unsatisfactory as it only indicates that the algorithm slows down. A more satisfactory indication of convergence is to check that the equation (\ref{eq:slackness}) is satisfied within a tolerance, since the equations (\ref{eq:alt_edge})  are satisfied exactly throughout the algorithm. However, this demands that the inverse $K=\Sigma^{-1}$ is calculated. 

This inversion may be computationally demanding, so we suggest that it  is only done when the algorithm is  slowing down. If the likelihood equations still are not fulfilled to the desired tolerance, we continue the iteration, but now we do not need a full inversion of $\Sigma$ at every step, as there is a simple procedure for updating $K$ after $\Sigma$ has been updated to $\tilde\Sigma=D_u\Sigma$. 
From (\ref{eq:update}) we have
\begin{equation}\label{eq:tildeSigma}\tilde\Sigma_{ru}=\tilde\Sigma_{ur}\transp=\Sigma_{rb}(\Sigma_{bb})^{-1}\Sigma_{bu}, \quad\tilde \Sigma_{bu}=\Sigma_{bu}=S_{bu}, \quad \tilde\Sigma_{uu}=\Sigma_{uu}=S_{uu}.\end{equation}
With a similar partitioning of $K=\Sigma^{-1}$ and $\tilde K=\tilde \Sigma^{-1}$ 
we get from (\ref{eq:tildeSigma}) and the fact that $\tilde \Sigma_{cc}=\Sigma_{cc}$ 
\begin{equation}\label{eq:k_identity}(\tilde K\tilde\Sigma)_{ru}=I_{ru}=0_{ru}= \tilde K_{rr}\Sigma_{rb}(\Sigma_{bb})^{-1}S_{bu}+\tilde K_{rb}S_{bu}+\tilde K_{ru}S_{uu},
\end{equation}
and similarly
$$0_{rb}=\tilde K_{rr}\Sigma_{rb}+\tilde K_{rb}\Sigma_{bb}+\tilde K_{ru}S_{ub}.$$ The last identity implies
$$\tilde K_{rr}\Sigma_{rb}=-\tilde K_{rb}\Sigma_{bb}-\tilde K_{ru}S_{ub},$$ which, 
inserted  into (\ref{eq:k_identity}) yields
\begin{eqnarray*}0_{ru}&=&
 -\tilde K_{rb}\Sigma_{bb}(\Sigma_{bb})^{-1}S_{bu}-\tilde K_{ru}S_{ub}(\Sigma_{bb})^{-1}S_{bu}+\tilde K_{rb}S_{bu}+\tilde K_{ru}S_{uu}
 \\&=&
\tilde K_{ru}\left\{S_{uu}-S_{ub}(\Sigma_{bb})^{-1}S_{bu }\right\}.\end{eqnarray*}
Since $S_{uu}-S_{ub}(\Sigma_{bb})^{-1}S_{bu }>0$, we conclude that $\tilde K_{ru}=0_{ru}$. 
This again implies
$$(\tilde\Sigma\tilde K)_{bu}=0_{bu}=\Sigma_{br}\tilde K_{ru}+\Sigma_{bb}\tilde K_{bu}+\Sigma_{bu}\tilde K_{uu}=
\Sigma_{bb}\tilde K_{bu}+\Sigma_{bu}\tilde K_{uu}$$
whence, if we as before let $\beta_{bu}=(\Sigma_{bb})^{-1}\Sigma_{bu}$, we have $$\tilde K_{bu}= -(\Sigma_{bb})^{-1}\Sigma_{bu}\tilde K_{uu}=-\beta_{bu}\tilde K_{uu}.$$ 
We further get $$1=\tilde\Sigma_{ur}\tilde K_{ru}+\Sigma_{ub}\tilde K_{ub}+\Sigma_{uu}\tilde K_{uu}=\Sigma_{ub}\tilde K_{ub}+\Sigma_{uu}\tilde K_{uu}= \Sigma_{uu}\tilde K_{uu}-\Sigma_{ub}(\Sigma_{bb})^{-1}\Sigma_{bu}\tilde K_{uu}
$$ and thus
$$\tilde K_{uu}=\{\Sigma_{uu}-\Sigma_{ub}(\Sigma_{bb})^{-1}\Sigma_{bu}\}^{-1}= (\Sigma_{uu}-\Sigma_{ub}\beta_{bu})^{-1}.$$
Summarizing the above findings gives the following equations for determining $\tilde K_{cu}$:
  \begin{gather}\label{eq:newtheta}
 \beta_{bu}=(\Sigma_{bb})^{-1}\Sigma_{bu},\quad\tilde K_{ru}=0,\\
 \tilde K_{bu}=-\beta_{bu}\tilde K_{uu}, \text{ where }\, \tilde K_{uu}=(\Sigma_{uu}-\Sigma_{ub}\beta_{bu})^{-1}\label{eq:newtheta3}.\end{gather}
 Standard results for block matrices yield that
 $$(\tilde \Sigma_{cc})^{-1}=(\Sigma_{cc})^{-1}=K_{cc}-K_{cu}K_{uc}/K_{uu}=\tilde K_{cc}-\tilde K_{cu}\tilde K_{uc}/\tilde K_{uu},$$whereby
 $$\tilde K_{cc}= K_{cc}-K_{cu}K_{uc}/K_{uu}+\tilde K_{cu}\tilde K_{uc}/\tilde K_{uu}.$$
 Exploiting that $\tilde K_{ru}=0$ further yields
 \begin{gather}\label{eq:K11_update1}
 \tilde K_{rr}= K_{rr}-K_{ru}(K_{ur}/K_{uu}),\\ \label{eq:K11_update2}\tilde K_{rb}= K_{rb}-K_{ru}(K_{ub}/K_{uu}),\\
 \tilde K_{bb}= K_{bb}-K_{bu}(K_{ub}/K_{uu})+\tilde K_{bu}(\tilde K_{ub}/\tilde K_{uu}).\label{eq:K11_update3}\end{gather}

 Combining equations (\ref{eq:newtheta})--(\ref{eq:K11_update3})  yields a procedure for updating $K$ at every subsequent step in the iteration without inverting $\tilde\Sigma$. The most expensive operation could be (\ref{eq:newtheta}) if $b=\bound(u)$ is large; but from (\ref{eq:update}) we observe  that the update of $\Sigma_{ru}$ may also be expressed as
$\tilde\Sigma_{ru}=\Sigma_{rb}\beta_{bu}$. Hence $\beta_{bu}$ has already been computed when updating $\Sigma$, so  the additional work involved when updating $K$ is not as computationally demanding as it could appear. 
\subsection{Computational savings}
\label{sec:effort_ncd}
Still, the update of $K$ will take some effort and also here a substantial saving may be obtained by ignoring unnecessary updates. Introduce the \emph{maximum column sum norm} of a matrix $\Delta=\{\Delta_{\alpha\beta}\}$ as 
\begin{equation}\label{eq:mnorm}
\mnorm{\Delta}_1=\max_{\beta}\sum_{\alpha}|\Delta_{\alpha\beta}|\end{equation} 
and let the algorithm terminate when $\mnorm{K(\graf)-K}_1<2\varepsilon/n$, since the gradient of the dual of the log-likelihood function is
$$\nabla_\Sigma\,\ell^*(\Sigma)=\nabla_\Sigma\,
\left\{ \frac{n}{2}\left(-\log \det \Sigma-d\right)\right\}=\frac{n}{2}\{K(\graf)-K\}.$$ This also corresponds to the equation (\ref{eq:slackness})  to be satisfied within that tolerance when properly scaled. Since
$$\mnorm{K(\graf)-K}_1= \max_{u\in V}\mnorm{K_{ru}}_1,$$  we may ignore the update corresponding to $u\in V$ if $\mnorm{K_{ru}}_1<2\varepsilon/n$ and terminate the algorithm when all updates are ignorable.

\subsection{Finding a feasible $K$}\label{sec:kfeasible}

The inverse  $K=\Sigma^{-1}$ obtained when the iteration terminates may not have exact zero values for non-edges, and some modification is necessary to find a feasible value $\check K\in \mathbb{S}^{d\times d}_{\prec}(\graf)$. For example, one might use the procedure described in Algorithm 17.1 of \cite{hastie:etal:16}, which  amounts to using only (\ref{eq:newtheta})--(\ref{eq:newtheta3}) in the final cycle, pretending that convergence has been achieved and therefore ignoring  (\ref{eq:K11_update1})--(\ref{eq:K11_update3}). This procedure does not ensure $\check K$ to be positive definite and if $\check K$ is determined in this way, the result will depend on the order in which the nodes $u\in V$ are visited. It seems difficult to control the outcome of this \emph{ad hoc} procedure and when the dimension $d$ was large, early experiments regularly encountered problems, whence it was abandoned. 

Since $K$ is available, it appears more direct to calculate $\check K$ from $K$ by  replacing elements $K_{uv}$ for $uv\notin E$ with zero, i.e.\ letting $\check K=K(\graf)$. This also does not ensure that $\check K$ is positive definite, but if the algorithm has been run until the equations  (\ref{eq:slackness}) are fulfilled up to a sufficiently small tolerance measured in a matrix norm \citep[p.\ 340--41]{horn:johnson:13}, it will be.  

To see this, we argue as follows.
First, to avoid scaling problems, we assume without loss of generality that $S$ has been scaled as a correlation matrix.  Let $\lambda_{\max}(A)$ and $\lambda_{\min}(A)$ denote the largest and smallest eigenvalue of a symmetric matrix $A$ and let $\Delta=K-\check K$, so $\check K= K-\Delta$. Then we have
\begin{eqnarray*}\lambda_{\min}(\check K)&=& \inf_{\{x\transp x=1\}}x\transp\! (K-\Delta)\,x\;\geq\; \inf_{\{x\transp x=1\}}x\transp\! K \,x-\sup_{\{x\transp x=1\}}x\transp\! \Delta\, x\\&=&
\lambda_{\min}(K)-\lambda_{\max}(\Delta)=\frac{1}{\lambda_{\max}(\Sigma)}-\lambda_{\max}(\Delta)\end{eqnarray*} and hence $\check K$ will be positive definite if
$\lambda_{\max}(\Delta)\lambda_{\max(\Sigma)}<1.$ 
Further, since it holds  throughout the iterative process that
$\trace(S)=\trace(\Sigma)=d$, we have  that $\lambda_{\max}(\Sigma)< d$. 
If $\mnorm{\cdot}$ is any matrix norm, for example the maximum column sum norm in (\ref{eq:mnorm}),   \citet[Theorem 5.6.9]{horn:johnson:13} implies
$$\lambda_{\max}(\Delta)\leq \mnorm{\Delta}.$$ Hence if we continue the iteration at least until $\mnorm{\Delta}<d^{-1}$,   $\check K$ is guaranteed to be positive definite.  

For
 any pair $(K, \Sigma)$ with  $K\in \mathbb{S}^{d\times d}_{\prec}(\graf)$ and $\Sigma$  positive definite and satisfying $\Sigma(\graf)=S(\graf)$,  the definition of the dual function yields
$$ \log\det K-\trace(KS)\leq -\log\det\Sigma-d$$ 
 with equality if and only if the pair $(K,\Sigma)$ is optimal; hence
\begin{equation}\label{eq:lbound}B= -\frac{n}{2}\left(\log\det\Sigma+d\right)\end{equation}  yields an upper bound on the log-likelihood function. Comparing the likelihood function for  $\check K$  with the upper bound from the final value of $\Sigma$, gives a certificate for the log-likelihood function to be close to its maximum;  the likelihood function is always at most $\gamma$ from its optimum value where $\gamma$ is the \emph{duality gap}
 $$\gamma= \frac{n}{2}\left\{\trace(\check{K}S)-\log\det (\check{K}\Sigma)-d\right\}.$$ 
In contrast to neighbourhood coordinate descent,  $K$ is feasible during the entire computational process under iterative proportional scaling; 
the price paid is that  there is no easy certificate available, since  $K^{-1}$ is not dually feasible unless the optimum has been reached exactly. 
\subsection{Finding a dually feasible starting value}\label{sec:feasible}
The dual algorithms  demand a dually feasible $\Sigma$ as a starting value to guarantee convergence, i.e.\ a positive definite matrix  $\Sigma$ satisfying $\Sigma(\graf)=S(\graf)$. The maximum likelihood estimator exists in the model if and only if such a feasible $\Sigma$ exists, \emph{cf.} Section~\ref{sec:dual}. If $S$ is positive definite, this is not an issue. But if $S$ is based on $n$   observations  and has $f=n-1$ degrees of freedom with $f<d$, some additional effort is needed. 
We may still make a factorization as in (\ref{eq:factordet})
$$\det \Sigma = \det \Sigma_{cc}\left\{\Sigma_{uu}-\Sigma_{uc}(\Sigma_{cc})^{-}\Sigma_{cu}\right\},$$
where now $(\Sigma_{cc})^{-}$ is any generalized inverse to $\Sigma_{cc}$.
We  again maximize the value of the Schur complement $(\tilde \Sigma/\Sigma_{cc})$, keeping $\Sigma_{cc}$ fixed. As before, this is maximized for $$\tilde\Sigma_{ru}=
\Sigma_{rb}(\Sigma_{bb})^{-}\Sigma_{bu}$$
and after the update,  the Schur complement becomes
\begin{equation}\label{eq:semi_schur}(\tilde \Sigma/\Sigma_{cc}) =\Sigma_{uu}-
\Sigma_{ub}(\Sigma_{bb})^{-}\Sigma_{bu},\end{equation}
but the determinant $\det\tilde\Sigma$ may still be zero. However, as we shall see below, if updates are made in a suitable order, the rank will increase at every step and eventually provide a dually feasible starting value.

 Assume  we have ordered the vertices $V=\{1,\ldots,d\}$ of $\graf$. Here and in the following we  let $[i]=\{1,\ldots, i\}$.  If the ordering satisfies 
\begin{equation}\label{eq:colourbound}\deg_{[i]}(i) <f\text{ for all $i=1,\ldots, d$,}\end{equation}
we say that the ordering is a \emph{reverse $f$-colouring sequence.}  Here
$\deg_A(u)=|\bound(u)\setminus A|$ is the number of neighbours of $u$ that are not in $A$.

The \emph{colouring number} $\col(\graf)$ of an undirected graph \citep{erdos:hajnal:66} is equal to the smallest number $k$ such that an ordering with $\max\deg_{[i]}(i)<k$ exists. 
The colouring number is different from  the chromatic number $\chi(\graf)$ but satisfies $\chi(\graf)\leq \col(\graf)$.
The number $\delta(\graf)=\col(\graf)-1$ is also known as the \emph{degeneracy} of the graph, and the \emph{$k$-core} of the graph is empty if and only if $\col(\graf)\leq k$; here the $k$-core of $\graf$ is what is left after vertices of degree less than $k$ are recursively removed.  In other words, we are able to find an ordering satisfying (\ref{eq:colourbound}) if and only if $\col(\graf)\leq f$.

There is a simple algorithm that finds such an ordering. This proceeds by repeatedly choosing a vertex of miminum degree as the next vertex in the ordering, then removing the vertex and its incident edges. The algorithm is termed the \emph{smallest-first algorithm} and  can be implemented in $O(|V|+|E|)$ time and space as described by \cite{matula:beck:83}; in fact, this reference describes a smallest-last algorithm using the reverse ordering, but this is immaterial.
Next, we introduce the concept of relative generic rank. 
\begin{definition}A symmetric $d\times d$ matrix $S$  is said to have \emph{generic rank $k$ relative to $A\subseteq V$} with $|V|=d$ if it has rank $k$ and every $k\times k$ principal submatrix of $S$ that includes all entries in $A$ has rank $k$.  For  $A=\emptyset$ we simply say that \emph{$S$ has generic rank $k$}. 
\end{definition}

Observe that with this definition, a $d\times d$ matrix $S$ with generic rank $d$ relative to $A$, is simply a matrix with full rank $d$, so then the set $A$ plays no role. 

Further, we need the following lemma, which establishes that the procedure for   updating the covariance  increases generic rank in the appropriate way. 


\begin{lemma}\label{lem:alt_update}Let $\graf=(V,E)$ be an undirected graph with $|V|=d$ and $\Sigma$  a positive semidefinite matrix with generic rank $|A|+k<d$ relative to $A\subseteq V$. Let $u\notin A$  with   $\deg_A(u)<k$. 
Further, let $\tilde\Sigma=D_u\Sigma$. Then it holds that $\tilde\Sigma(\graf)=\Sigma(\graf)$, and $\tilde\Sigma$ has generic rank $|A|+k+1$ relative to $A\cup\{u\}$. 
\end{lemma}
\begin{proof}It is obvious that $D_u\Sigma(\graf)=\Sigma(\graf)$ as the update only changes elements of $\Sigma$ corresponding to non-edges.

Since $\deg_A(u)<k$ and $\Sigma$ has  generic rank $|A|+k$ relative to $A$,  $\Sigma_{A\cup b\cup\{u\},A\cup b\cup\{u\}}$ is positive definite. Hence $\Sigma_{ b\cup\{u\},b\cup\{u\}}$ is also positive definite  so combining with (\ref{eq:semi_schur}) for $c=V\setminus\{u\}$ we get  that
\begin{equation}\label{eq:new_update_det}(\tilde\Sigma/\Sigma_{cc})=\Sigma_{uu}-
\Sigma_{ub}(\Sigma_{bb})^{-}\Sigma_{bu}
=\Sigma_{uu}-\Sigma_{u b}(\Sigma_{bb})^{-1}\Sigma_{bu } >0\end{equation} and therefore, by rank additivity of the Schur complement \citep{guttman:46}
$$\rank(\tilde\Sigma)=\rank(D_u \Sigma)=\rank(\tilde\Sigma/\Sigma_{cc})+\rank(\Sigma_{cc})=1+(|A|+k).$$
We need to show that the rank of $\tilde\Sigma$ is generic relative to $A\cup \{u\}$; i.e.\ that  any  principal submatrix of $\tilde\Sigma$ with dimenstion $(|A|+k+1)\times (|A|+k+1)$ that contains $A$ and $u$ is positive definite. 
So assume $a$ satisfies $A\subseteq a\subseteq V\setminus\{u\}$ with $|a|=|A|+k$; 
the task is to show that  the matrix $$B=\tilde\Sigma_{a\cup u,a\cup u}=
\begin{pmatrix}\Sigma_{aa}&\tilde\Sigma_{au}\\ \tilde\Sigma_{u a}&\Sigma_{uu}\end{pmatrix} $$
is positive definite, since $B$ is  an arbitrary $(|A|+k+1)\times(|A|+k+1)$ principal submatrix of $\tilde\Sigma$ containing $A$ and $u$. 
Using Schur complements, we get 
$$\det B=\det\Sigma_{aa}(\Sigma_{uu}-\tilde\Sigma_{ua}\Sigma_{aa}^{-1}\tilde\Sigma_{au}).$$
The first factor $\det\Sigma_{aa}$ is positive because $\Sigma$  has generic rank $|A|+k$  relative to $A\subseteq a$; the second factor is
the residual variance after regressing $u$ onto the variables in $a$ when $\tilde\Sigma$ is the covariance, since $B$ is then the covariance  of the variables in $a\cup\{u\}$. This is at least as large as the residual variance after regressing $u$ onto all of  the variables in $c=V\setminus\{u\}$ and this is positive by (\ref{eq:new_update_det}); hence the second factor is also positive. 
We conclude that $\det B>0$ and $B$ is positive definite, as required. 
\end{proof}

Observe that $\tilde\Sigma$ does not have generic rank $|A|+k+1$ relative to $A$ only, since entries not involving $u$ have not been changed  and therefore any $(|A|+k+1)\times (|A|+k+1)$ principal submatrix not involving $u$ still has  rank $|A|+k$ and not $|A|+k+1$.

For $i=1,\ldots, d$ we let  $D_{[i]}=(D_i\cdots D_1)$; repeated use of Lemma~\ref{lem:alt_update} now yields:
\begin{proposition}\label{prop:alt_update}Let $\graf=([d],E)$ be an undirected graph with $\col(\graf)\leq k$ with vertices numbered as a reverse $k$-colouring sequence. Further, let $\Sigma$ be a positive semidefinite $d\times d$ matrix with generic rank $k$. Then it holds for $i=1,\ldots d$  that $D_{[i]}\Sigma(\graf) =\Sigma(\graf)$, and $D_{[i]}\Sigma$ has generic rank $\min(k+i,d)$ relative to $[i]$. 
\end{proposition}
\begin{proof} The proof is by induction after $i$. For $i=1$ this is Lemma~\ref{lem:alt_update} since $\deg(1)<k$ because the vertices are ordered as a reverse $k$-colouring sequence. 

So assume the statement has been established for $i\leq m$ and let $i=m+1$.  
The inductive assumption implies that $D_{[m]}\Sigma(\graf)=\Sigma(\graf)$ and $D_{[m]}\Sigma$ has generic rank $\min(m+k,d)$ relative to $[m]$. Since the sequence is a reverse $k$-colouring sequence, $\deg_{[m+1]}(m+1)<k$ and therefore 
$$|[m+1]\cup \bound_{[m+1]}(m+1)|\leq  m+k.$$
So if we let $c^m=[m+1]\cup \bound_{[m+1]}(m+1)$, we have that
$(D_{[m]}\Sigma)_{c^m,c^m}$ has full rank and therefore is positive definite. Since we have $\bound(m+1)\subseteq c^m$, this  also holds for the principal submatrix $(D_{[m]}\Sigma)_{\clos(m+1),\clos(m+1)}$.  
Noting that $D_{[m+1]}=D_{m+1}D_{[m]}$, Lemma~\ref{lem:alt_update} yields that 
$$D_{[m+1]}\Sigma (\graf)=D_{[m]}\Sigma(\graf)=\Sigma(\graf)$$ so $D_{[m+1]}\Sigma $ fits on the diagonal and edges of $\graf$; further, $D_{[m+1]}\Sigma$ has generic rank $\min(k+m+1,d)$ relative to $[m+1]$. This completes the proof.
\end{proof}

The following corollary yields a simple method for finding a starting value  when the graph $\graf$ has low colouring number.
\begin{corollary}\label{cor:smart_start}Let $S$ be a positive semidefinite $d\times d$ matrix 
with generic rank $f$, let $\graf=([d],E)$ be an undirected graph with $\col(\graf)\leq f$, 
and assume the numbering of vertices represents  a reverse $f$-colouring sequence.
Then
$\Sigma^0 =D_{[d]}S$ is positive definite and 
$\Sigma^0(\graf)=S(\graf)$, hence $\Sigma^0$ is dually feasible.
\end{corollary}
\begin{proof} This is a rephrasing of Proposition~\ref{prop:alt_update} for $i=d$.
%
%
\end{proof}

As a consequence, we obtain the following:
\begin{corollary}\label{cor:existence} If $\col(\graf)\leq f$, the maximum likelihood estimator of $\Sigma$ in the Gaussian graphical model with graph $\graf$ based on an empirical covariance matrix $S$ with  $f$ degrees of freedom exists with probability one. 
\end{corollary}
\begin{proof}An empirical covariance matrix $S$ as above has generic rank $f$ with probability one if the distribution of observations has density with respect to Lebesgue measure on $\mathbb{R}^d$ \citep{eaton:perlman:73}. Corollary~\ref{cor:smart_start} then yields a dually feasible starting value.  The result follows.  \end{proof}

Observe that already $\Sigma^0=\Sigma_{[d-f]}S$ is dually feasible, so we only need to make as many updates as the rank deficiency $d-f$ to obtain a dually feasible starting value. Also,  if the updates during neighbourhood coordinate descent are made according to a reverse $f$-colouring sequence, the algorithm converges from the starting value $\Sigma^0=S$, since the update $\Sigma$ will be positive definite after the first round of iterations.

Corollary~\ref{cor:existence} is a reformulation of Theorem~3.5 in \cite{gross:sullivant:18}, since $\col(\graf)\leq f$ if and only if the $f$-core of $\graf$ is empty. The condition is also equivalent to the upper bound for the Gaussian rank of $\graf$ given in Theorem~1.1 of \cite{bendavid:15}, expressed in terms of graph degeneracy $\delta(\graf)$ since $\col(\graf)=\delta(\graf)+1$.
 The condition $\col(\graf)\leq f$ is not sufficient for existence of the maximum likelihood estimator and may be weakened \citep{bernstein:etal:22}, but it is an easy and effective condition to check, whereas a simple necessary and sufficient condition is still not known. 

If the graph is sparse so that the number of vertices of degree less than $f$ is larger than the rank deficiency $d-f$ of the empirical covariance matrix, Lemma~\ref{lem:alt_update} implies that we may obtain a starting value without reordering the variables:
\begin{proposition}\label{prop:quick_start} Let $S$ be a positive semidefinite $d\times d$ matrix 
with generic rank $f$, let $\graf=([d],E)$ be an undirected graph    and  let $V'=\{v\in V\cd  \deg(v)<f\}$. If $|V'|=d'\geq d-f$ we let $D_{V'}=\prod_{v\in V'}D_v$, where the composition is made in any order. 
Then
$\Sigma^0 =D_{V'}S$ is positive definite and 
$\Sigma^0(\graf)=S(\graf)$, hence $\Sigma^0$ is dually feasible.
\end{proposition}
\begin{proof}This follows from Corollary~\ref{cor:smart_start} by observing that any ordering satisfying $V'=[d']$ represents a reverse $f$-colouring sequence.
\end{proof}
And further, as a direct consequence, no starting value is needed for the algorithm to converge if the maximal degree of the graph is lower than the degrees of freedom for the empirical covariance matrix:
\begin{corollary}\label{cor:no_start} Let $S$ be a positive semidefinite $d\times d$ matrix 
with generic rank $f$, let $\graf=([d],E)$ be an undirected graph with $\max_{v\in V}\deg(v)<f$. Then the neighbourhood coordinate descent algorithm started at $S$ converges to the maximum likelihood estimator of $\Sigma$ based on $S$.
\end{corollary}
\begin{proof}Follows directly from Proposition~\ref{prop:quick_start} with $V'=[d]$.
\end{proof}

\section{Empirical study}
\label{sec:study}

\subsection{Implementation of the algorithms}
The algorithms have been implemented in R \citep[version 4.3.2]{r}.  
We have made an implementation
based on
C++ using the \pkg{RcppArmadillo} package 
\citep[version 0.12.6.6.0]{rcpparmadillo:14}.  The experiments have been run on AMD EPYC 7302 16-core processors with 64 CPUs and 3 GHz clock frequency.
The implementation is naive in the sense
that we store the full matrices $K$ and not just the non-zero elements. On the other hand, for iterative proportional scaling we  store and calculate $S_{cc}$ and its inverse $(S_{cc})^{-1}$ for all relevant subsets $c$ once and for all to use for updating $K$ in (\ref{eq:kupdate1})
so the empirical covariance matrix $S$ is itself not needed.  The code producing the results in this section as well as an .html file with a more detailed output is available as supplementary material from \texttt{https://github.com/hojsgaard/gRips}.

The algorithms were applied  to data representing a sample of 102 instances of the expression of 6033 genes associated with prostate cancer, originating from \cite{prostate_data} and published in the R package \pkg{spls} \citep{spls}. In addition,  artificial samples were produced with 102 instances of the relevant number of variables, all entries simulated from the standard normal $N(0,1)$ distribution. 
A default tolerance of $\varepsilon=10^{-3}$ was used throughout. 

For iterative proportional scaling, the algorithms were run until the likelihood equations were satisfied with  an error less that $\varepsilon'=2\varepsilon/102=10^{-3}/51$, when measured by the maximum absolute deviation norm $\norm{\Sigma(\graf)-S(\graf)}_\infty$, to match the tolerance to the gradient of the log-likelihood function.  For the covariance based algorithm, updates were ignored if  $\norm{\Sigma_{cc}-S_{cc}}_\infty <\varepsilon'$, as described in Section~\ref{sec:effort_ips}.

Similarly,  neighbourhood coordinate descent 
 was first run until consequtive  values of $\Sigma$ differed by less than $\varepsilon'$ when measured by the maximum column sum norm (\ref{eq:mnorm}). 
Then a second series of cycles was  run, updating $K$ alongside $\Sigma$ until 
$\mnorm{K(\graf)-K}_1\leq \varepsilon''$, where $\varepsilon''= \min(\varepsilon',d^{-1})$ as described in
Section~\ref{sec:kfeasible}, thereby ensuring that $\check K=K(\graf)$ was positive definite. 
In this second cycle, updates of any vertex $u\in V$ was ignored if $\mnorm{K_{ru}}_1<\varepsilon''$ as described at the end of Section~\ref{sec:effort_ncd}.  All iterations were performed according to a smallest-first ordering of the variables so no specific starting values were needed. In the actual experiments, all graphs  had a maximal degree well below 100,  so it was in fact  unnecessary to update via a specific ordering.  This could potentially have reduced computing time by a small amount, but we chose to include the reordering phase in the timings.
%

\subsection{Comparing the algorithms for moderate size dense graphs}
\label{sec:increase-dens}

We first investigated the computing time for the iterative proportional scaling algorithms and neighbourhood coordinate descent  for random graphs with 100 variables of varying density. Model fitting for the scaling algorithms
was based on edges or cliques.  
The median computing times in seconds over five random graphs are displayed in Table~\ref{tab:increase-dens}. 

\begin{table}[htb]
\def~{\hphantom{0}}
\tbl{
    Median computing time in seconds over five random graphs on 100 vertices for the covariance (COV) and concentration (CON) based iterative proportional scaling algorithms, applied  cliquewise or edgewise, and for  neighbourhood coordinate descent (NCD)}{%
\begin{tabular}{cccccccccccc}
\\
&&\multicolumn{5}{c}{Simulated data}&\multicolumn{5}{c}{Prostate data}\\
&Expected&\multicolumn{2}{c}{Cliquewise}&\multicolumn{2}{c}{Edgewise}&&\multicolumn{2}{c}{Cliquewise}&\multicolumn{2}{c}{Edgewise}&\\
Density&\# of  edges&CON&COV&CON&COV&NCD&CON&COV&CON&COV&NCD\\
10\%&495&0.2&0.03&0.6&0.04&0.05&2&0.1&7&0.2&0.1\\
30\%&1,485&1&0.1&3&0.1&0.08&13&0.4&82&3&0.1\\
50\%&2,475&14&1&10&0.3&0.1&47&3&1,607&19&0.3\\
70\%&3,465&299&24&48&1&0.2&576&34&20,323&227&0.3
\end{tabular}}
\label{tab:increase-dens}
\end{table}

The general picture  in Table~\ref{tab:increase-dens} is that the covariance based version is considerably faster than the concentration based algorithm  and the effect is stronger when fitting a 
dense model and when models are updated edgewise.
There are several reasons for this: Firstly, for sparse graphs, relatively many cliques would be pairs and hence there is little difference between edgewise or cliquewise updating.
Secondly, when the
model is dense, the cliques will be relatively large so updating
$\Sigma$ as in (\ref{eq:sigmaupdate}) 
will be
time consuming. Thirdly,  a random dense graph will typically have many
large cliques sharing many variables. This means that the same edges
are updated several times during each iteration. 
Finally, the speedup due to ignoring inefficient local updates has smaller effect when using cliques rather than edges.

The neighbourhood coordinate descent algorithm is obviously very fast for this type of models and is less sensitive to the density or sparsity of the graph. It fits the densest model for the prostate data in less than a second. 

Computing times are systematically shorter when the algorithms are applied to simulated data. This is most likely a reflection of the fact that the empirical covariance matrix would tend to fit any of the models investigated and therefore be closer to the final estimate from the outset, hence demanding fewer iterations in the fitting procedures.
%
For the densest model, the concentration based scaling algorithm sometimes failed to reach convergence within the specified limit when applied edgewise to the prostate data, even after 50,000 iterations. Hence the concentration based algorithm seems unfeasible when applied edgewise to dense graphs. 

We also observe that the algorithms in this and similar cases slow down much before they stop, while the likelihood function is still far from the correct value. This highlights the danger using slowness as convergence criterion.

\subsection{Comparison with neighbourhood coordinate descent for large, sparse graphs}
\label{sec:comp-with-graph}

Next we  shall compare computing times for the edgewise covariance based scaling algorithm and neighbourhood coordinate descent for larger dimensions.
We use the same data as in the previous section for comparison. 

For random graphs with more than 100 vertices, experiments may become complicated as there is a risk that the maximum likelihood estimate does not exist. So to extend the above comparisons to larger scale and a higher degree of sparsity without this difficulty, we first investigate the behaviour for rectangular grids. For such grids, the maximum likelihood estimate exists with probability one for a sample covariance matrix with just three degrees of freedom since $\col(\graf)=3$ for any rectangular grid. This follows since  there is always a corner with only two neighbours that may be removed in the smallest-first algorithm. 
Hence 102 observations are plenty. 
For a rectangular grid, there is no difference between updating edgewise or cliquewise, as all cliques consist of exactly one edge.  Computing times for the algorithms and various grid sizes  are displayed in Table~\ref{tab:gridexp}. 
\begin{table}[htb]
\def~{\hphantom{0}}
\tbl{
    Computing time in seconds for covariance based  scaling (COV) and neighbourhood coordinate descent (NCD) over a rectangular grid}{%
\begin{tabular}{cccccccc}\\
&&&&\multicolumn{2}{c}{Simulated data}&\multicolumn{2}{c}{Prostate data}\\
Grid size&\# of variables&\# of edges&Density&COV&NCD&COV&NCD\\
$20\times 25$&500&955&0.8\%&0.4&1&1&2\\
$40\times 25$ &1,000&1,935&0.4\% &3  &5&7 &10\\
$40\times 50$& 2,000&3,910&0.2\%&18&29&57&64\\
$80\times 50$&4,000&7,870&0.1\%&169&220&500&379
\end{tabular}}
\label{tab:gridexp}
\end{table}

Table~\ref{tab:gridexp} indicates that the covariance based scaling algorithm is comparable to neighbourhood coordinate descent at this level of sparsity; it fits the model of an $80\times50$ grid to the prostate data in about eight minutes and to  simulated data in three minutes, whereas neighbourhood coordinate descent  uses about six minutes for the prostate data and about four minutes for simulated data.

Our final experiments are similar to those in \cite{xu:etal:15}. 
We consider random graphs that are constructed by adding random edges  to a random tree with probabilities 0.001, 0.005, and 0.010 respectively; the number of variables varying from 500 to 4,000, and up to 6,000 for the sparsest case. Computing times are displayed in Table~\ref{tab:treeexp}. 
\begin{table}[htb]
\def~{\hphantom{0}}
\tbl{
    Median computing time in seconds over five random trees with additional edges for covariance based iterative proportional scaling (COV) and neighbourhood coordinate descent (NCD)}{%
\begin{tabular}{cccccccc}\\ 
&&&\multicolumn{2}{c}{Simulated data}&\multicolumn{2}{c}{Prostate data}\\
Density&\#  variables&Exp. \# of  edges&COV&NCD&COV&NCD\\
\multirow{5}{*}{0.001}&500&623&0.2&1&0.3&2\\
&1,000&1,498&1&6&4&13\\
&2,000&3,996&22&65&334&149\\
&4,000&11,993&782&612&5,975&911\\
&6,000&23,990&3,364&1,615&25,749&2,096\\
&&&&&&\\
\multirow{4}{*}{0.005}&500&1,120&0.4&1&1&2\\
&1,000&3,491&3&6&89&16\\
&2,000&11,984&266&72&5,331&268\\
&4,000&43,969&5,124&578&75,500&1,209\\
&&&&&&\\
\multirow{4}{*}{0.010}&500&1,742&0.9&1.3&5&2\\
&1,000&5,984&28&9&463&25\\
&2,000&21,969&869&75&18,685&498\\
&4,000&83,939&10,500&679&242,290&2,955\\
\end{tabular}}
\label{tab:treeexp}
\end{table}

 The computing times of the scaling algorithm is comparable to neighbourhood coordinate descent  when the graph is sparse and the size is moderate, whereas the latter is considerably faster in higher dimension and for higher densities.  Neighbourhood coordinate descent is generally less sensitive to sparsity of the graph. Also, the computing times for the algorithms as before are much higher for the prostate data than for the simulated data.  
 
A few computations were also made using neighbourhood coordinate descent for random trees with additional edges added with  density  0.05. Already at dimension 3,000, the colouring number of the random graphs exceeded the number of observations  so the MLE would not  exist unless  the number of observations was increased to around 130. With more than 200,000 edges, iterative proportional scaling was unfeasible. Using neighbourhood coordinate descent on the basis of 130 simulated observations yielded total fitting times around 500 seconds for 3,000 variables. 

\cite{xu:etal:15} report best computing times for  iterative proportional scaling with 4,000 variables and simulated data to be 1,044 and 2,407 seconds for densities  0.005 and 0.010 respectively, which should be compared to our 4,682  and 9,681 seconds. However, these numbers are not quite comparable since we have used a much stricter convergence criterion. We made experiments with a weaker criterion, but abandoned them; although the convergence was faster, the accuracy for the maximized likelihood function was not satisfactory. 

It is possible that a  partitioning of the edges as described in \cite{xu:etal:15} might potentially also  speed up the scaling algorithm. However, it is not clear to us whether the overhead of finding this partition and setting up the corresponding structure is included in the CPU times reported there; since this involves simulated annealing, it might be a considerable bottleneck.   When \cite{xu:etal:15} report results on real data with more than 6000 variables; the model has first been determined with the graphical lasso so tends to fit better than a random graph and as we have seen, this has a strong effect on the computing times. Also, the models used are extremely sparse, with a total of 11,000 to 15,000 edges and a maximal connected component between 4,000 and 5,000 variables. They then report computing times up  to 6,000 seconds, which should be compared to our 738 seconds  using scaling and 133 seconds using neighbourhood coordinate descent for a random tree with 4,000 variables and random edges added at density 0.001, since this has a similar number of edges.

\section{Discussion}
\label{sec:discussion}
We have described a  version of iterative proportional
scaling  for fitting Gaussian graphical models avoiding the NP-complete task of identifying the cliques and updating the concentration and covariance matrices simultationeously without inverting large matrices.   The increase of speed  is in particular noticeable when graphs are sparse.
 
Further, we have described a version of neighbourhood coordinate descent which is extremely fast, scales well with the size of the graph, is less sensitive to the density of the graph, and is guaranteed to provide a positive definite concentration matrix with zeros in the right places for sparse graphs. In addition, the algorithm provides a certificate that guarantees the accuracy of the value of the likelihood function.  
As a consequence, we recommend to use neighbourhood coordinate descent by default.


\begin{thebibliography}{}

\bibitem[Banerjee et~al., 2008]{banerjee:etal:08}
Banerjee, O., El~Ghaoui, L., and d'Aspremont, A. (2008).
\newblock Model selection through sparse maximum likelihood estimation for
  multivariate {G}aussian or binary data.
\newblock {\em Journal of Machine Learning Research}, 9:485--516.

\bibitem[Ben-David, 2015]{bendavid:15}
Ben-David, E. (2015).
\newblock Sharp lower and upper bounds for the {G}aussian rank of a graph.
\newblock {\em Journal of Multivariate Analysis}, 139:207--218.

\bibitem[Bernstein et~al., 2022]{bernstein:etal:22}
Bernstein, D.~A., Dewar, S., Gortler, S.~J., Nixon, A., Sitharam, M., and
  Theran, L. (2022).
\newblock Computing maximum likelihood thresholds using graph rigidity.
\newblock arXiv:2210.11081.

\bibitem[Boyd and Vandenberghe, 2004]{boyd:vandenberghe:04}
Boyd, S. and Vandenberghe, L. (2004).
\newblock {\em Convex Optimization}.
\newblock Cambridge University Press, Cambridge, UK.

\bibitem[Chung et~al., 2019]{spls}
Chung, D., Chun, H., and Keles, S. (2019).
\newblock {\em \pkg{spls}: Sparse Partial Least Squares (SPLS) Regression and
  Classification}.
\newblock R package version 2.2-3.

\bibitem[Eaton and Perlman, 1973]{eaton:perlman:73}
Eaton, M.~L. and Perlman, M.~D. (1973).
\newblock The non-singularity of generalized sample covariance matrices.
\newblock {\em The Annals of Statistics}, 1:710--717.

\bibitem[Eddelbuettel and Sanderson, 2014]{rcpparmadillo:14}
Eddelbuettel, D. and Sanderson, C. (2014).
\newblock \pkg{Rcpp{A}rmadillo}: Accelerating {R} with high-performance {C}++
  linear algebra.
\newblock {\em Computational Statistics and Data Analysis}, 71:1054--1063.

\bibitem[Erd{\"o}s and Hajnal, 1966]{erdos:hajnal:66}
Erd{\"o}s, P. and Hajnal, A. (1966).
\newblock On chromatic number of graphs and set-systems.
\newblock {\em Acta Mathematica Academiae Scientarum Hungarica}, 17:61--99.

\bibitem[Friedman et~al., 2008]{friedman:etal:08}
Friedman, J., Hastie, T., and Tibshirani, R. (2008).
\newblock {Sparse inverse covariance estimation with the graphical lasso.}
\newblock {\em Biostatistics}, 9:432--441.

\bibitem[Friedman et~al., 2019]{glasso:19}
Friedman, J., Hastie, T., and Tibshirani, R. (2019).
\newblock {\em {\sc glasso}: Graphical Lasso for Estimation of Gaussian
  Graphical Models}.
\newblock R package version 1.11.

\bibitem[Grone et~al., 1984]{grone:etal:84}
Grone, R., Johnson, C.~R., {de S{\'a}}, E.~M., and Wolkowicz, H. (1984).
\newblock Positive definite completions of partial {H}ermitian matrices.
\newblock {\em Linear Algebra and its Applications}, 58:109--124.

\bibitem[Gross and Sullivant, 2018]{gross:sullivant:18}
Gross, E. and Sullivant, S. (2018).
\newblock The maximum likelihood threshold of a graph.
\newblock {\em Bernoulli}, 24:386--407.

\bibitem[Guttman, 1946]{guttman:46}
Guttman, L. (1946).
\newblock Enlargement methods for computing the inverse matrix.
\newblock {\em The Annals of Mathematical Statistics}, 17:336--343.

\bibitem[Hara and Takemura, 2010]{hara:takemura:10}
Hara, H. and Takemura, A. (2010).
\newblock A localization approach to improve iterative proportional scaling in
  {G}aussian graphical models.
\newblock {\em Communications in Statistics---Theory and Methods},
  39:1643--1654.

\bibitem[Hastie et~al., 2016]{hastie:etal:16}
Hastie, T., Tibshirani, R., and Friedman, J. (2016).
\newblock {\em The Elements of Statistical Learning}.
\newblock Springer-Verlag, New York, 2nd edition.
\newblock 11th printing.

\bibitem[Horn and Johnson, 2013]{horn:johnson:13}
Horn, R.~A. and Johnson, C.~R. (2013).
\newblock {\em Matrix Analysis}.
\newblock Cambridge University Press, Cambridge, UK, 2nd edition.

\bibitem[Jirou{\accent 7 s}ek and P{\accent 7 r}eu{\accent 7 c}il,
  1995]{jirousek:preucil:95}
Jirou{\accent 7 s}ek, R. and P{\accent 7 r}eu{\accent 7 c}il, R. (1995).
\newblock On the effective implementation of the iterative proportional fitting
  procedure.
\newblock {\em Computational Statistics and Data Analysis}, 19:177--189.

\bibitem[Lauritzen and Zwiernik, 2022]{lauritzen:zwiernik:22}
Lauritzen, S. and Zwiernik, P. (2022).
\newblock {Locally associated graphical models and mixed convex exponential
  families}.
\newblock {\em The Annals of Statistics}, 50:3009--3038.

\bibitem[Lauritzen, 1996]{lauritzen:96}
Lauritzen, S.~L. (1996).
\newblock {\em Graphical Models}.
\newblock Clarendon Press, Oxford, United Kingdom.

\bibitem[Marchetti et~al., 2020]{marchetti:etal:20}
Marchetti, G.~M., Drton, M., and Sadeghi, K. (2020).
\newblock {\em \textsc{ggm}: Graphical Markov Models with Mixed Graphs}.
\newblock R package version 2.5.

\bibitem[Matula and Beck, 1983]{matula:beck:83}
Matula, D.~W. and Beck, L.~L. (1983).
\newblock Smallest-last ordering and clustering and graph coloring algorithms.
\newblock {\em Journal of the Association for Computing Machinery},
  30:417--427.

\bibitem[{R Core Team}, 2023]{r}
{R Core Team} (2023).
\newblock {\em R: A Language and Environment for Statistical Computing}.
\newblock R Foundation for Statistical Computing, Vienna, Austria.

\bibitem[Singh et~al., 2002]{prostate_data}
Singh, D., Febbo, P., Ross, K., Jackson, D., Manola, J., Ladd, C., Tamayo, P.,
  Renshaw, A., {DA}mico, A., Richie, J., Lander, E., Loda, M., Kantoff, P.,
  Golub, T., and Sellers, W. (2002).
\newblock Gene expression correlates of clinical prostate cancer behavior.
\newblock {\em Cancer Cell}, 1:203--209.

\bibitem[Speed and Kiiveri, 1986]{speed:kiiveri:86}
Speed, T.~P. and Kiiveri, H. (1986).
\newblock {G}aussian {M}arkov distributions over finite graphs.
\newblock {\em The Annals of Statistics}, 14:138--150.

\bibitem[Uhler, 2019]{uhler:19}
Uhler, C. (2019).
\newblock Gaussian graphical models.
\newblock In Maathuis, M., Drton, M., Lauritzen, S., and Wainwright, M.,
  editors, {\em Handbook of Graphical Models}, pages 217--238. CRC Press, Boca
  Raton, FL.

\bibitem[Wermuth and Scheidt, 1977]{wermuth:scheidt:77}
Wermuth, N. and Scheidt, E. (1977).
\newblock Algorithm {AS} 105: Fitting a covariance selection model to a matrix.
\newblock {\em Journal of the Royal Statistical Society. Series C (Applied
  Statistics)}, 26:88--92.

\bibitem[Whittaker, 1990]{whittaker:90}
Whittaker, J. (1990).
\newblock {\em Graphical Models in Applied Multivariate Statistics}.
\newblock John Wiley and Sons, Chichester.

\bibitem[Xu et~al., 2011]{xu:etal:11}
Xu, P.-F., Guo, J., and He, X. (2011).
\newblock An improved iterative proportional scaling procedure for {G}aussian
  graphical models.
\newblock {\em Journal of Computational and Graphical Statistics}, 20:417--431.

\bibitem[Xu et~al., 2012]{xu:etal:12}
Xu, P.-F., Guo, J., and Tang, M.-L. (2012).
\newblock An improved {H}ara--{T}akamura procedure by sharing computations on
  junction tree in {G}aussian graphical models.
\newblock {\em Statistics and Computing}, 22:1125--1133.

\bibitem[Xu et~al., 2015]{xu:etal:15}
Xu, P.-F., Guo, J., and Tang, M.-L. (2015).
\newblock A localized implementation of the iterative proportional scaling
  procedure for {G}aussian graphical models.
\newblock {\em Journal of Computational and Graphical Statistics}, 24:205--229.

\bibitem[Yuan and Lin, 2007]{yuan:lin:07}
Yuan, M. and Lin, Y. (2007).
\newblock Model selection and estimation in the {G}aussian graphical model.
\newblock {\em Biometrika}, 94:19--35.

\end{thebibliography}
\end{document}